\documentclass[preprint2,a4paper]{aastex}
\input{crab.defs}
\usepackage{epsf}
\usepackage{emulateapj5}
\usepackage{onecolfloat5}
\shortauthors{Bhat et al.}
\shorttitle{Bright Giant Pulses}

\def\SM{{\rm SM}}

\newcommand{\Wtau}{\mbox{$W_{\tau}$}}

\newcommand{\be}{\begin{eqnarray}}
\newcommand{\ee}{\end{eqnarray}}

\begin{document}
\twocolumn[
\title{Bright Giant Pulses from the Crab Nebula Pulsar: Statistical Properties,
Pulse Broadening and Scattering due to the Nebula}
\medskip
\author{N. D. Ramesh Bhat$^1$, Steven J. Tingay$^2$, and Haydon S. Knight$^1$}
\affil{$^1$Centre for Astrophysics \& Supercomputing, Swinburne University, Hawthorn, Victoria 3122, Australia \\ 
$^2$Department of Imaging and Applied Physics, Curtin University, Bentley, Western Australia, Australia
}
\begin{abstract}
We report observations of Crab giant pulses made with the Australia Telescope Compact Array 
and a baseband recorder system, made simultaneously at two frequencies, 1300 and 1470 MHz.
These observations were sensitive to pulses with amplitudes \ga 3 kJy and widths \ga 0.5 $\mu$s.
Our analysis led to the detection of more than 700 such bright giant pulses over 3 hours,
and using this large sample we investigate their amplitude, width, arrival time and 
energy distributions.
The brightest pulse detected in our data has a peak amplitude of $\sim$ 45 kJy and a width 
of $\sim$ 0.5 $\mu$s, and therefore an inferred brightness temperature of $\sim 10^{35}$ K.
The duration of giant-pulse emission is typically $\sim$1 $\mu$s, however it can 
also be as long as 10 $\mu$s. 
The pulse shape at a high time resolution (128 ns) shows rich diversity and complexity in 
structure and is marked by an unusually low degree of scattering. 
We discuss possible implications for scattering due to the nebula, and for underlying 
structures and electron densities.
\end{abstract}
\keywords{pulsars: general -- pulsars: individual (Crab pulsar)
-- ISM: structure -- ISM: individual (Crab Nebula) -- scattering}
]


\section{Introduction} \label{s:intro}

The Crab Nebula pulsar B0531+21 is well known for its emission of giant radio pulses, and was 
originally discovered through the detection of such pulses \citep{staelin1968}. 
These sporadic, large-amplitude and short-duration bursts can be hundreds or thousands of times 
more energetic than regular pulses \citep[e.g.][]{lundgrenetal1995}, and they remain one of the 
most enigmatic aspects of pulsar radio emission.
While several other pulsars are now known to emit giant pulses 
\citep[e.g.][]{cognardetal1996,johnstonetal2004,knightetal2006}, only 3 objects, {\it viz.}
the Crab and the millisecond pulsars B1937+21 and J1823--3021A, generate giant pulses numerous 
enough to allow detailed studies of their characteristics. 

Observations so far have unravelled several fundamental properties of giant pulse emission. 
It is now fairly well established that the fluctuations in their amplitudes are due to changes 
in the coherence of the radio emission \citep{lundgrenetal1995}, and that they are 
superpositions of extremely narrow nanosecond structures \citep{hankinsetal2003}. 
Observations also suggest that the emission is broadband, extending over several hundreds 
of MHz \citep{sallmenetal1999,popovetal2006}. Another important characteristic of giant pulses 
is their tendency to originate in a very narrow phase window of regular radio emission; and 
for the Crab, these windows even coincide with the phases of the high-energy (from infrared 
to $\gamma$-ray) emission \citep{moffett-hankins1996}. 
Finally, the distribution of giant pulse energies is known to follow a power-law form 
\citep{argyle-gower1972,lundgrenetal1995}, much in contrast to the Gaussian or exponential 
distribution typical of regular pulses \citep[e.g.][]{hesse-wielebinski1974}. 

Of all the giant-pulse--emitting pulsars, the Crab is the most well-studied. It is known to 
emit pulses as energetic as $\sim 10^4$ times regular pulses \citep{cordesetal2004} and 
shows structures persisting down to 2 ns, with inferred brightness temperatures as high 
as $10^{37}$ K \citep{hankinsetal2003}. 
Given such extreme 
short durations, giant pulses are best studied using data from baseband observations. Such 
data allow coherent dedispersion of voltage samples to remove the deleterious effect of 
interstellar dispersion, thereby yielding more accurate descriptions of pulse structure, 
shape and amplitude. However, the bulk of the observational studies of the Crab until the early 
2000s were carried out using traditional filterbank or spectrometer data, due to limitations 
of data throughput and computing. 
With the advent of wide-bandwidth recorders and affordable high-power computing, 
these limitations are being gradually overcome. As a result, baseband observations are 
increasingly employed in giant pulse studies \citep[e.g.][]{soglasnovetal2004,popov-stappers2007}. 

In this paper, we report our observations of Crab giant pulses made using the Australia Telescope 
Compact Array (ATCA) and the baseband recorder recently developed for the Australian Long 
Baseline Array.
We detected over 700 giant pulses with amplitudes \ga 3 kJy and 
widths $\sim$0.5 to 10 $\mu$s, and using this large sample we investigate aspects such as
the pulse amplitude, width, arrival time and energy distributions, as well as details of 
the pulse structure and shape. Our observations show a much lower degree of scattering 
than reported before and we discuss possible implications for scattering due to the nebula.

\section{Observations and Data Processing} \label{s:obsproc}

\subsection{Observations with the ATCA} \label{s:obs}

\begin{figure}[t]
\epsscale{1.0}
\plotone{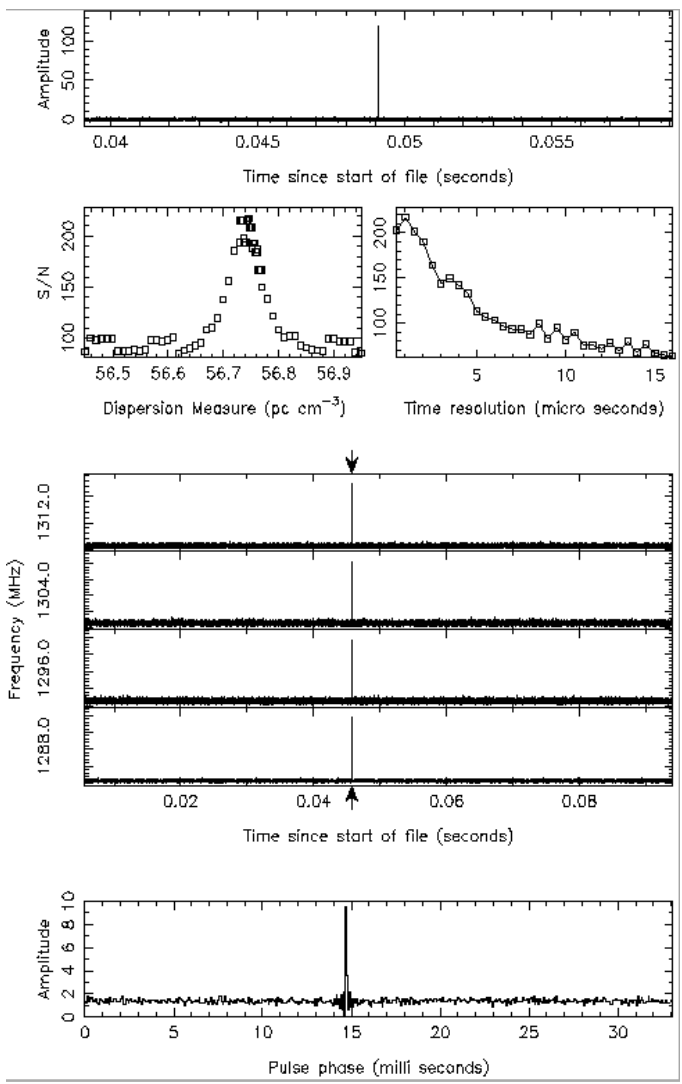}
\caption{Example plots illustrating our giant pulse detection. 
{\it Top}: Dedispersed time series of a short data segment around the giant pulse.
{\it Middle panels}: Diagnostic plots from the giant pulse search; 
plots of signal-to-noise ratio vs trial dispersion measure and time resolution, and 
dedispersed pulses over four 8-MHz sub-bands of the 32 MHz recorder bandwidth.  
{\it Bottom}: The dedispersed pulse over a time window of 33 ms (i.e. one rotation 
period). 
}
\label{fig:gpsearch}
\end{figure}

Observations of the Crab pulsar were made in January 2006.
Although the array offers maximal sensitivity for pulsar observations in its {\it tied-array} 
mode, our observing set-up was designed also to meet certain technical goals in addition to 
pulsar observations, and offered a sensitivity equivalent to that of a single ATCA antenna.
In this set-up, the array was configured to record two independent IF (intermediate
frequency) signals, each 32 MHz in 
bandwidth, from a single antenna. 
At the antenna, two 64-MHz bands were Nyquist-sampled at 128 MHz, with four bits per voltage sample, 
for two orthogonal linear polarisations.
These data were converted to circular polarisations by inserting a $90^{\circ}$ phase shifter in 
the signal path. 
The signals were then digitally filtered to provide two-bit sampled voltage time series for the 
32 MHz bandwidth.
This results in an aggregate data rate of 512 megabits per second.
The system provided flexibility to record either two separate frequency bands from a single 
antenna, or two identical frequency bands from two different antennas. 
The pulsar observations were made over a total duration of 3 hours; the first half of which used
two 32-MHz bands (dual polarisation) centred at frequencies 1300 and 1470 MHz, while the remaining
half used two identical
frequency bands centred at 1300 MHz from two different antennas. 
This accumulated a total data volume of 5.5 terabytes, but split into a large number of 
short blocks of 10 s duration each to facilitate further analysis and processing.
The data were stored onto disks for offline processing and all processing operations, such as 
coherent dedispersion, detection and search for giant pulses, were performed at the Swinburne
University of Technology Supercomputing Facility.

\subsection{Data Decoding and Dedispersion} \label{s:encod}

Data samples from both the polarisation channels of the two IF bands were organised into a single
data stream while recording, and were subsequently decoded into separate IF data streams before 
de-dispersion and further processing.
As our processing software is designed to handle a single data stream at a given time, this meant 
two separate passes for processing each short block of data. 
The unpacked data from this stage were then dedispersed using the coherent dispersion removal 
technique originally developed by \citet{hankins1971}. 
The procedures we have adopted are similar to those described in \citet{knightetal2006} and
\citet{bhatetal2007}. 
Voltage samples are first Fourier transformed to the frequency domain and the spectra 
are divided into a series of sub-bands. 
Each sub-band is then multiplied by an inverse response filter (kernel) for the ISM 
\citep[e.g.][]{hankins-rickett1975,vanstraten2003}, and then Fourier transformed back 
to the time domain to construct a time series with a time resolution coarser than the
original data. 
By splitting the input signal into several sub-bands (four in our case), the dispersive 
smearing is essentially reduced to that of an individual sub-band. 
This also means the procedure uses shorter transforms than single-channel coherent 
dedispersion and is consequently a more computationally efficient method.
The coherent filterbank stream data obtained in this manner are then square-law detected, 
corrected for dispersive delays between the sub-bands, and summed in polarisation 
to construct a single coherently dedispersed time series for the entire band.

\subsection{Search for Giant Pulses} \label{s:search}

Following the procedures of decoding and dedispersion, we performed a rigorous
search for giant pulses within each 10 s block of data.
Our pulse detection procedure involved progressive smoothing of time series 
with matched filters of widths ranging from 0.5 to 16 $\mu$s 
in steps of 0.5 $\mu$s, and identifying the intensity samples that exceed a 
set threshold (e.g. 12$\sigma$ for the 0.5 $\mu$s smoothing time). 
In addition to performing dedispersion at the Crab's nominal DM of 56.75 \dmu, 
we also performed this procedure over a large number of adjacent DM values 
(typically over a DM range $\approx$ 0.5 \dmu, in steps of 0.001 \dmu).
For each DM and the matched filter width, we computed the signal-to-noise ratio 
(S/N) of the pulse amplitude over a short stretch of data centred at the 
pulse maximum.
From this analysis, diagnostic plots are generated for each candidate giant 
pulse as shown in Fig.~\ref{fig:gpsearch}. 
These plots are subjected to a careful human scrutiny to 
discriminate real giant pulses from spurious signals.

\subsection{Summary of Detections} \label{s:det}

We adopted a rather conservative threshold of 12$\sigma$ in order to account for possible
departures of noise statistics from those expected of pure white noise and also to limit 
the number of candidate signals to a reasonable number.
With this threshold and the 0.5 $\mu$s final time resolution ($\Delta t $) employed in our 
analysis, a pulse will need to be $\ga$3 kJy in amplitude to enable a clear detection (see 
\S~\ref{s:sens}).
We detected 706 giant pulses from our 3-hr long observation, of which 413 are from 
the first half of observations that  were made simultaneously at 1300 and 1470 MHz. 
However, only 70\% of these 413 were detected in both frequency bands.
This fraction is similar to that reported by \citet{sallmenetal1999} from their 
observations at 600 and 1400 MHz, although quite different fractions have been 
reported from observations at widely separated frequencies \citep{kostyuketal2003} 
and at lower frequencies \citep{popovetal2006}. 
This probably suggests that not all pulses are entirely broadband, a characteristic 
presumably intrinsic to the giant-pulse emission phenomenon. 
Our sample includes pulses with widths ranging from 0.5 to 10 $\mu$s.
Given $\Delta t $ = 0.5 $\mu$s, any shorter-duration pulses that may be present in our 
data will be smoothed to 
an effective resolution\footnote{In practice, pulses that are narrower than 0.5 $\mu$s 
will be detected as either 0.5 $\mu$s wide (most flux in one sample) or 1 $\mu$s wide 
(pulse comes in just on the border of two samples) in our analysis.} of 0.5 or 1 $\mu$s.
Moreover, pulses broader than 10 $\mu$s are likely to be weaker than our detection threshold. 
The strongest pulse in our observations is 0.5 $\mu$s in duration and has a S/N of 217. 

\subsection{System Sensitivity and Flux Calibration} \label{s:sens}

The Crab Nebula is a fairly bright and extended source in the radio sky, with a flux density of 
$\sim$955$\,\nu^{-0.27}$ Jy (\citet{bietenholzetal1997}; where $\nu$ is the frequency in GHz) 
and a characteristic diameter of $\sim$5$^{\prime}$.5.
Thus, in general, there will be a significant contribution from the nebula to the system 
noise ($S$), depending on the frequency of observation and the coverage of the 
nebula within the telescope beam.
However, in our case, the problem is much simplified as the nebula is unresolved by a single 
antenna of the array (half-power beam width $\approx$ $33^{\prime}$ at our observing frequencies). 
In fact, the entire nebular region occupies only a small fraction ($\approx$ 3\%) of the beam 
solid 
angle.
Measurements made in parallel with the observations yield system temperatures (\Tsys) of 
$114 \, \pm \, 3$ and $103 \, \pm \, 2$ K respectively at 1300 and 1470 MHz. 
Thus, assuming a nominal gain (G) of $\approx$ 0.1 \kpjy for a single ATCA antenna at L-band, 
these measurements translate to system equivalent flux densities (\Ssys=\Tsys/G) of 1140 
and 1030 Jy respectively at 1300 and 1470 MHz.
Scaling for our processing parameters and the recording bandwidth ($\Delta B$), and also 
accounting for the loss of S/N due to our 2-bit digitisation,
these estimates will correspond to system noise of 250 and 227 Jy at the 1-$\sigma$ 
level\footnote{$ S=\eta N_{\sigma} \Tsys G^{-1} (n_{pol} \Delta B \Delta t)^{-1/2} $, where
$\eta$ is the loss of S/N due to digitisation, $ N _{\sigma} $ is the detection threshold in 
units of $\sigma$, $ n_{pol} $ is the number of polarisations summed.}.
In other words, minimum detectable pulse amplitudes of 3 and 2.7 kJy respectively at 1300 
and 1470 MHz for a 12-$\sigma$ threshold. 

\section{Statistical Properties of Giant Pulses} \label{s:stat}

\begin{figure}[t]
\epsscale{0.8}
\plotone{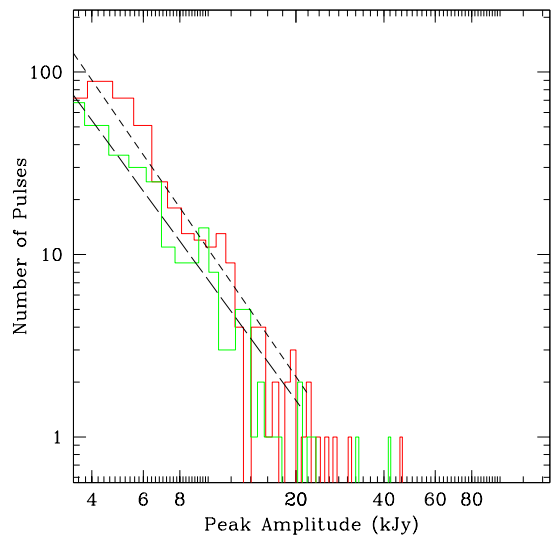}
\caption{Histograms of giant-pulse amplitudes at 1300 and 1470 MHz (in red and green 
respectively) constructed from our sample of giant pulses. 
Estimates for the best fit slope ($\beta$) are $-2.33\pm0.14$ and $-2.20\pm0.18$ at
1300 and 1470 MHz respectively, and are shown as short-dashed and long-dashed lines. 
The brightest pulse detected in our data (at 1300 MHz) has a peak amplitude of 
$\sim$45 kJy.
}
\label{fig:amp}
\end{figure}

\subsection{Pulse Amplitudes and Energies} \label{s:energy}

An important distinguishing characteristic of giant pulses (GPs) is their power-law 
distribution of amplitudes and energies.
The amplitude determination can be potentially influenced by factors such as
smearing due to instrument and any residual dispersion, as well as external
effects such as scattering and scintillation due to the ISM. 
As a result, oftentimes measurements tend to underestimate the true amplitudes. 
Observations of \citet{lundgrenetal1995} and \citet{cordesetal2004} (at 430 MHz) were marked by 
significant instrumental and dispersion smearing (315 and 152 $\mu$s respectively) and were 
more prone to scintillation and scattering given their observing frequencies \la 1 GHz.  
On the other hand, smearing due to instrument or residual dispersion is negligibly small for 
baseband observations, and consequently our data enable a better and more accurate 
characterisation of the pulse amplitudes. 
Fig.~\ref{fig:amp} shows the distributions of pulse amplitudes for our data.  

\begin{figure}[t]
\epsscale{1.0}
\plotone{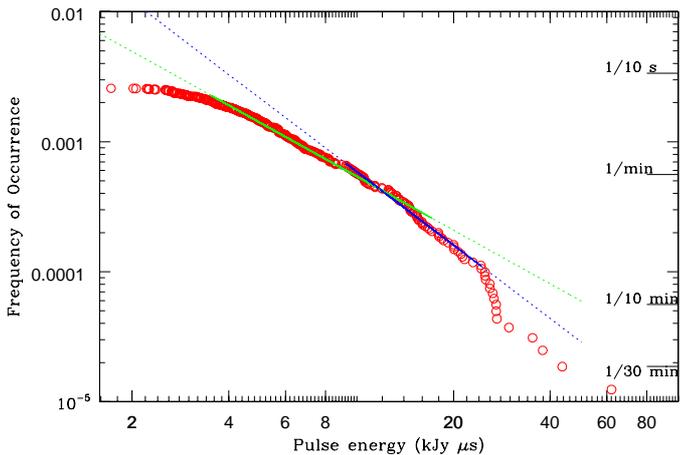}
\caption{Cumulative distribution of giant-pulse energies at 1300 MHz.
The slope estimates of the linear segments below and above an apparent break near 
$\sim$10 kJy $\mu$s are --1.4 and --1.9 respectively, while the overall slope is 
--1.6.}
\label{fig:cdp}
\end{figure}

The cumulative energy distribution provides a meaningful way of characterizing the frequency of 
occurrence of GPs. 
Following \citet{knightetal2006}, we define this distribution in terms of the probability of a 
pulse having energy greater than \Eo, and can be expressed as 
\be
P ( E > \Eo) = K \, E _0 ^{\alpha}
\ee
Figure~\ref{fig:cdp} shows such a distribution for our data. 
The pulse energy is estimated by integrating the amplitude bins over the extent of emission.
The uncertainties in the energy estimates depend on the pulse strength and width, and may 
range from $\sim$0.5\% for strong and narrow pulses, to as much as $\sim$35\% for weak, 
broad pulses. 
Such large uncertainties at low energies, and possible modulations in pulse amplitudes due
to scintillation, may probably explain a gradual flattening seen at low energies.
In any case, much in agreement with the earlier work, no evidence is seen for a high-energy 
cut-off, suggesting that exceedingly bright and energetic pulses are potentially observable 
over longer durations of observation.

Excluding the segments near the low and high ends of the distribution where the behaviour 
tends to depart from a power-law, we obtain a best-fit value of $-1.59 \, \pm \, 0.01$ 
for $\alpha$ (where the error is purely formal) and $K$ = 3.97.
Interestingly, this slope is comparable to the published values for other GP-emitters such 
as PSRs J0218+4232 \citep{knightetal2006}, B0540--69 \citep{johnstonetal2004} and B1937+21 
\citep{cognardetal1996,soglasnovetal2004}.
 
However, it is known from the work of \citet{moffett1997} that a single slope does not 
accurately describe the energy distribution at 1400 MHz and that there occurs a break 
around an energy of $\sim$2 kJy $\mu$s, where the slope $\alpha$ changes from $-3$ to $-1.8$.
Given that this energy corresponds to the detection threshold of our GP search (see \S~\ref{s:sens}), 
our slope estimate of $-1.6$ is in general agreement with Moffett's work. 
Recent work of \citet{popov-stappers2007} suggests that the slope estimate depends on the 
pulse width, evolving from $-1.7$ to $-3.2$ when going from their shortest (4 $\mu$s) to 
longest (65 $\mu$s) GPs.
Their analysis also shows a clear evidence for a break where the power-law tends to steepen
(their Fig. 2).
Such a break is also apparent in our data, as the slope tends to steepen near $\sim$10 kJy $\mu$s.
Adopting this as the break point, we obtain slope estimates of $-1.37 \, \pm \, 0.01$ and $-1.88 \, \pm \, 0.02$ 
for the segments below and above it (K = 3.72 and 4.42 respectively).
  
The pulse energy distribution also serves as a useful guide for estimating the rates of 
occurrence of giant pulses. 
For instance, on the basis of our observations, a pulse that is about 50 times more 
energetic than regular pulses\footnote{Assuming mean pulse energies estimated from
parameters in the published literature \citep{manchesteretal2005}.} can be expected 
roughly once in 8500 pulse periods, i.e., approximately once every 5 minutes.
The most energetic pulse detected in our data has an estimated energy of 66 kJy $\mu$s; 
i.e. an energy per $\mu$s that is almost $\sim 10^5$ times larger than that of regular 
pulses.

The brightest pulse detected in our data has an estimated peak flux density (\Snu) of $\sim$ 45 kJy 
and an effective width of 0.5 $\mu$s. 
The equivalent brightness temperature (\Tb) is given by (based on the light-travel size and ignoring 
relativistic dilation), 
\be
\Tb = \left( { \Snu \over 2 \, \kB } \right) \left( { D \over \nu \, \delt } \right)^2,
\ee
where $ \nu $ is the frequency of observation, 
\delt~is the pulse width, \kB~is the Boltzmann constant and D is the Earth-pulsar distance. 
The inferred brightness temperature for our strongest pulse is therefore $\sim10^{35}$ K.
This is almost $\sim$1000 times brighter than the brightest pulse detected in the Arecibo 
observations of \citet{cordesetal2004}.
To the best of our knowledge, this marks the brightest pulse ever recorded from the Crab pulsar
within the L-band frequency range (i.e. $\sim$1--2 GHz). 
Detection of such excessively bright giant pulses offers a promising technique for finding pulsars 
in external galaxies \citep{johnston-romani2003,cordesetal2004}.

\begin{figure}[b]
\epsscale{1.0}
\plotone{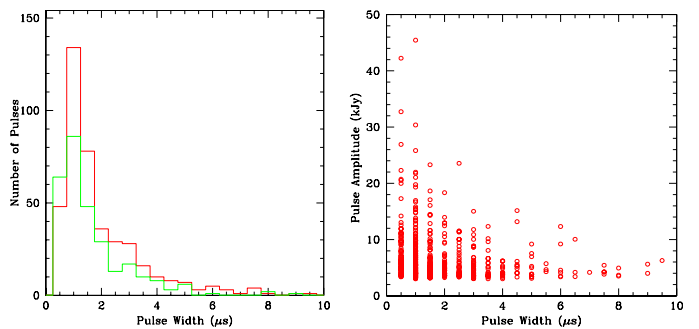}
\caption{{\it Left}: 
Distributions of giant-pulse widths at 1300 and 1470 MHz (in red and green respectively). 
{\it Right}: Plot of pulse amplitudes vs the effective pulse widths. 
An apparent low cut-off at 0.5 $\mu$s and a discretization in pulse width are due to the
0.5 $\mu$s final time resolution adopted for our processing.
}
\label{fig:wid}
\end{figure}

\subsection{Pulse Widths} \label{s:wid}

There have been very few attempts of characterising giant-pulse widths.
The first analysis by \citet{lundgrenetal1995} was limited by their coarse time sampling 
(205 $\mu$s) and was further constrained by a severe dispersion smearing (70 $\mu$s).
More recent work of \citet{popov-stappers2007} had a much higher time resolution 
(4.1 $\mu$s).
As our observations were made at a higher frequency than the above (hence the effect of 
scattering is reduced) and a final time resolution (0.5 $\mu$s) that is eight-fold better than 
\citet{popov-stappers2007}, our analysis allows a more accurate characterization 
of giant-pulse widths.

Our observations show that many giant pulses tend to have a significant structure 
at 0.5 $\mu$s resolution, ranging from a simple narrow spike to several closely 
spaced components within a few $\mu$s. 
Following \citet{lundgrenetal1995} and \citet{popov-stappers2007}, we adopt 
the notion of ``effective pulse width'' \we, which is essentially the averaging time that yields 
the maximum S/N.
In most cases, it is a close representation of true pulse width, although it may be slightly 
biased to stronger component(s) in the case of highly structured pulses.
The histogram of widths obtained in this manner is shown in Fig.~\ref{fig:wid}.
The distribution is highly skewed, with measured widths ranging from 0.5 to 10 $\mu$s and 
with a clear peak at 1 $\mu$s. 
Giant pulses of widths larger than 10 $\mu$s are absent in our data; however, this may be a 
selection effect as such pulses may very well be below our detection threshold.
Our analysis confirms a general tendency for stronger pulses to be narrower, an observation
also noted by \citet{popov-stappers2007}. 
 
Our distribution of giant-pulse widths can be compared with a similar plot  obtained for 
GPs from the millisecond pulsar PSR B1937+21 by \citet{soglasnovetal2004}. 
Similarities in the two distributions are quite striking, despite the fact that GPs from PSR B1937+21 
are intrinsically narrower (\we \la 15 ns) and show little structure. 
Thus, it appears that exponential-tailed distributions are probably common characteristics of 
giant-pulse widths. Studies of more pulsars are necessary to confirm such a conjecture. 

\begin{figure}[b]
\epsscale{1.0}
\plotone{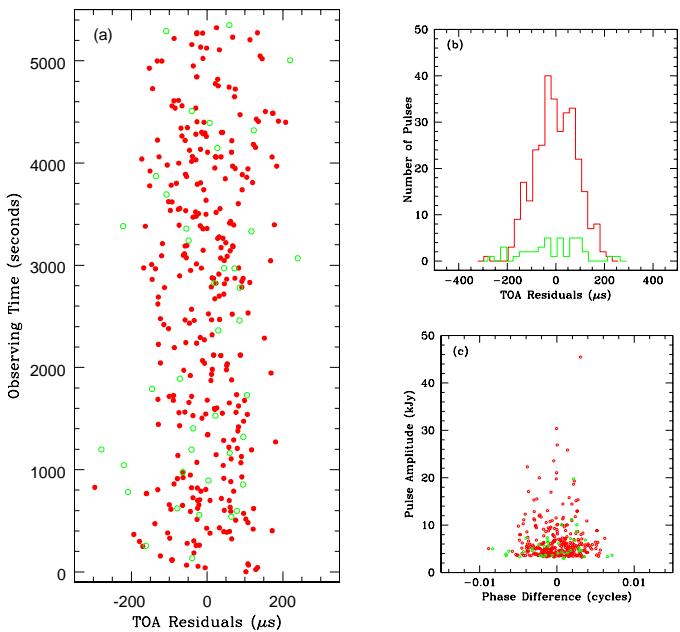}
\caption{Arrival times of giant pulses detected in the main pulse and interpulse 
regions (the red and green symbols respectively) at 1300 MHz ({\it left}), along 
with their statistics ({\it right, top}) and joint statistics with pulse amplitudes 
({\it right, bottom}). 
}
\label{fig:toas}
\end{figure}

\subsection{Arrival Times and Phases} \label{s:toas}

One of the most striking properties of Crab GPs is their occurrence within well-defined, 
relatively narrow longitude ranges of regular radio emission.
This is also true with the millisecond pulsars PSRs B1937+21 and 
J0218+4232, except that in the case of the former, GPs tend to occur near the trailing edges
of the main pulse and interpulse \citep{soglasnovetal2004}, and in the latter case, 
they occur near the two minima of the integrated pulse profile \citep{knightetal2006}. 
In order to determine the pulse arrival times, we use the pulsar's spin-down model along 
with the {\tt TEMPO} software package to obtain the anticipated pulse arrival phases. 
As most pulses in our data are very narrow, our data allow precise determinations of 
the arrival times and phases. 
Pulse times-of-arrival (TOAs) obtained in this manner are plotted in Fig.~\ref{fig:toas}.

The total range of longitudes where GPs occur (in the main pulse region) is approximately 
$\pm$ 200 $\mu$s, or $\pm 2^{\circ}.2$ in angular units, with an RMS of 84 $\mu$s ($0^{\circ}.9$). 
Thus, a vast majority of GPs (75\%) occur within a narrower window of $\pm 100$ $\mu$s.
As contributions from the dispersive and instrumental smearing are negligible in our case, 
this RMS corresponds essentially to the intrinsic pulse-phase jitter and can be 
compared to a value of 90 $\mu$s estimated by \citet{lundgrenetal1995} for their data at 800 MHz. 
The plot of joint statistics of timing residuals and pulse amplitudes shows no obvious correlation 
except for a general tendency for stronger pulses to originate within narrower phase windows.
A similar property was also noted by \citet{cordesetal2004} in their Arecibo observations at a 
frequency of 430 MHz. 
Finally, a vast majority of GPs tend to occur within the main pulse window $-$ 87\% at the main 
pulse region and the remainder 13\% in the interpulse region.
This is in excellent agreement with the published values in the literature
(\citet{popov-stappers2007}: 84\% and 16\% respectively at 1200 MHz; 
\citet{kostyuketal2003}: 86\% and 14\% respectively at 2228 MHz).

\section{Pulse Broadening and Scattering due to the Nebular Plasma} \label{s:scatt}

\begin{figure}[t]
\epsscale{1.0}
\plotone{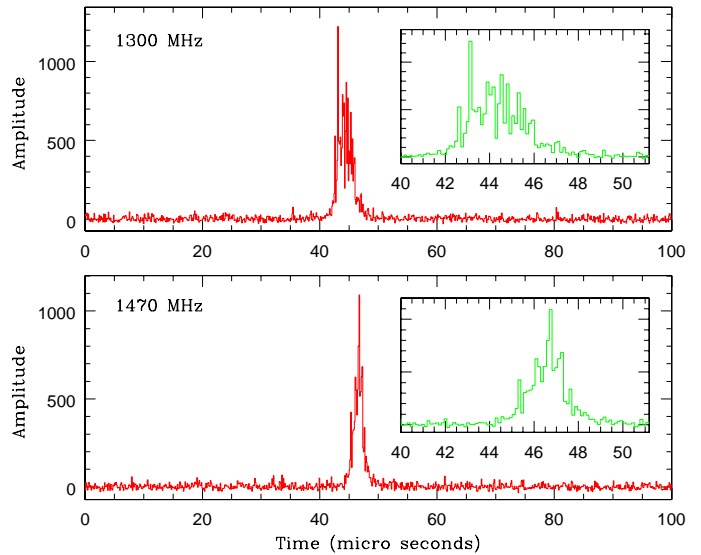}
\caption{A bright giant pulse at 1300 and 1470 MHz, where the plotted time window corresponds 
to $\approx$0.003 cycles of the pulsar's rotation period. 
This pulse pair is displayed at a time resolution of 128 ns and illustrates significant 
differences seen in pulse width and structural details between the two frequencies. 
}
\label{fig:gps}
\end{figure}

\subsection{Pulse Shape and Estimation of Scattering} \label{s:shape}

Figure \ref{fig:gps} shows an example of a bright giant pulse from our data.
At a time resolution of 128 ns, the pulse is resolved into multiple narrow components, 
and this readily confirms the basic picture of GPs comprising fine structure 
on very short timescales. 
Discerning such fine structure is however limited by time resolution and the 
smearing due to multipath scattering. 
This pulse pair also exemplifies the differences seen in pulse structure between 1300 
and 1470 MHz, a detailed analysis of which is beyond the scope of the present work. 
Our pulse shapes can be compared to observations of \citet{sallmenetal1999} $\sim$10 yr ago 
(their Fig. 1), and it is striking that our observations are marked by a much lower degree of 
scattering.

\begin{figure}[t]
\epsscale{1.0}
\plotone{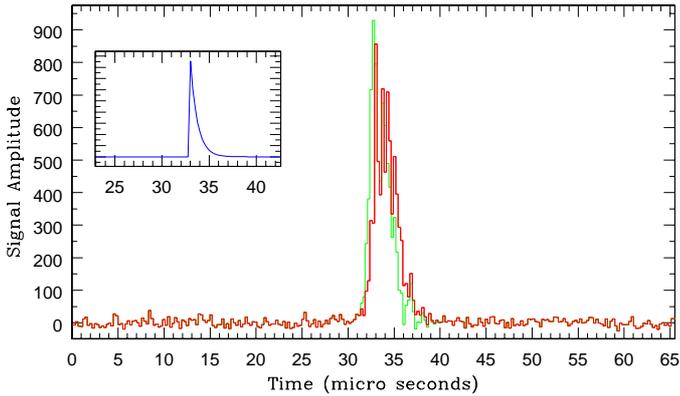}
\caption{A bright giant pulse from observations at 1300 MHz. 
The time window corresponds to 0.002 cycles of the pulse period.
The reconstructed pulse (green) and the best-fit pulse-broadening function (inset) 
from our deconvolution procedure are also shown. 
The $e^{-1}$ point of this one-sided exponential gives an estimate of the 
pulse-broadening time, for which our measured value is $0.8\pm0.4$ $\mu$s.
}
\label{fig:clean}
\end{figure}

In order to estimate the pulse broadening due to interstellar 
scattering \citep[e.g.][]{williamson1972,cordes-lazio2001}, we 
adopt the CLEAN-based deconvolution approach developed by \citet{bhatetal2003}.
Unlike the traditional frequency-extrapolation approach \citep[e.g.][]{lohmeretal2001,kuzminetal2002}, this 
method makes no prior assumption of the intrinsic pulse shape, and thus offers a more robust 
means of determining the underlying pulse broadening function (PBF).
The procedure involves deconvolving the measured pulse shape in a manner quite similar to 
the CLEAN algorithm used in synthesis imaging, while searching for the best-fit PBF and 
recovering the intrinsic pulse shape. 
It relies on a set of figures of merit that are defined in terms of positivity and symmetry 
of the resultant deconvolved pulse and some parameters characterizing the noise statistics 
in order to determine the best-fit PBF. 
For the purpose of our analysis, we assume the simplest and most commonly used form for the 
PBF that corresponds to a thin-slab scattering screen geometry. 
The functional form for such a PBF is a one-sided exponential \citep[e.g.][]{williamson1972} 
and is given by
\be
G(t) = \left ( { 1 \over \tau _d } \right) \, \exp \left( { -t \over \taud } \right) \, U(t),
\ee
where $U(t)$ is the unit step function, $U(<0) = 0, U(\ge 0) = 1$.
Fig.~\ref{fig:clean} shows the best-fit PBF and the reconstructed pulse obtained in this manner.
The $ { \rm e^{-1} } $ point of this PBF is the scattering time \taud, for which our deconvolution 
procedure yields an estimate of $0.8\pm0.4$ $\mu$s.

This measurement of a low scattering is further supported by the following observational facts.
First, a pulse broadening of $\sim$0.8 $\mu$s will imply rapid intensity decorrelations in
frequency on characteristic scales $ \sim 1/2 \pi \taud$ $\sim$ 0.1 to 0.3 MHz and this is 
confirmed by our analysis.
Second, as seen in Fig.~\ref{fig:wid}, a large number of pulses detected in our data have widths 
$\sim$1 $\mu$s, just about what we would expect given the $\sim$0.8 $\mu$s broadening 
due to scattering and the 0.5 $\mu$s time resolution.

\subsection{Scattering due to the Nebula} \label{s:nebula}

Our measurement of \taud can be compared to that reported by \citet{sallmenetal1999} based on 
their observations in 1996.
Their value of $95\pm5$ $\mu$s at 600 MHz would scale to 4.3 $\mu$s at 1300 MHz, assuming a 
canonical $\nu^{-4}$ dependence expected from scattering due to a turbulent plasma screen.
This is five times larger than our measurement.
A somewhat smaller value is expected on the basis of observations of \citet{kuzminetal2002} 
in the year 2000. 
However, our measurement is consistent with that of \citet{bhatetal2007} at 200 MHz, if we 
extrapolate their measurements using their revised frequency scaling of $\nu^{-3.5\pm0.2}$. 

As described in \S~\ref{s:search}, our GP detection procedure includes determination of the 
best DM by performing dedispersion over many trial values around its nominal value. 
As most pulses in our data are very narrow, this procedure allows a precise determination of the 
Crab's true DM at our observing epoch. 
An estimate of $56.751\pm0.001$ \dmu obtained in this manner is further confirmed by measuring the 
time delay between the pulse arrival times at 1300 and 1470 MHz. 
This value is in excellent agreement with that reported in the Jodrell Bank (JB) monthly 
ephemeris
\citep{lyne1982} 
on the nearest date of our observing, but it is significantly lower than that 
measured near the observing epochs of \citet{sallmenetal1999}. 

Thus, over a 10-yr time span between Sallmen et al's and our observations, the pulsar's 
DM decreased by 0.09 \dmu while the pulse broadening decreased by a factor of 5. 
Given the strong frequency dependence of pulse broadening ($ \taud \propto \nu^{-x}$) and the 
observational evidence for the scaling index $x$ changing over the time \citep{kuzminetal2002,bhatetal2007}, 
it is more meaningful to adopt a 
frequency independent parameter such as the scattering measure (SM) for comparison purposes.
This parameter quantifies the total scattering along the line of sight (LOS) and is defined 
as the LOS integral of \cn, which is the spectral coefficient of the wavenumber spectrum of electron 
density irregularities. It can be related to \taud via the relation
$\taud \approx 1.1\, \Wtau~\SM^{6/5}\nu^{-x}D$, where $\nu$ is in GHz, D is in kpc, 
and $\Wtau$ is a geometric factor that depends on the LOS-distribution of scattering 
material \citep{cordes-rickett1998}. 
Assuming $\Wtau=1$, we can estimate the {\it effective} SM for a uniform medium, and 
for the measurements of Sallmen et al's and ours, we obtain 
values of $1.3 \times 10^{-2}$ and $3 \times 10^{-3}$ \smu respectively.
That is, the SM changed by 0.01 \smu when the DM changed by 0.09 \dmu, or a factor 4 
decrease in the scattering strength associated with a DM change of only 0.16\%. 

\begin{figure}[t]
\epsscale{1.0}
\plotone{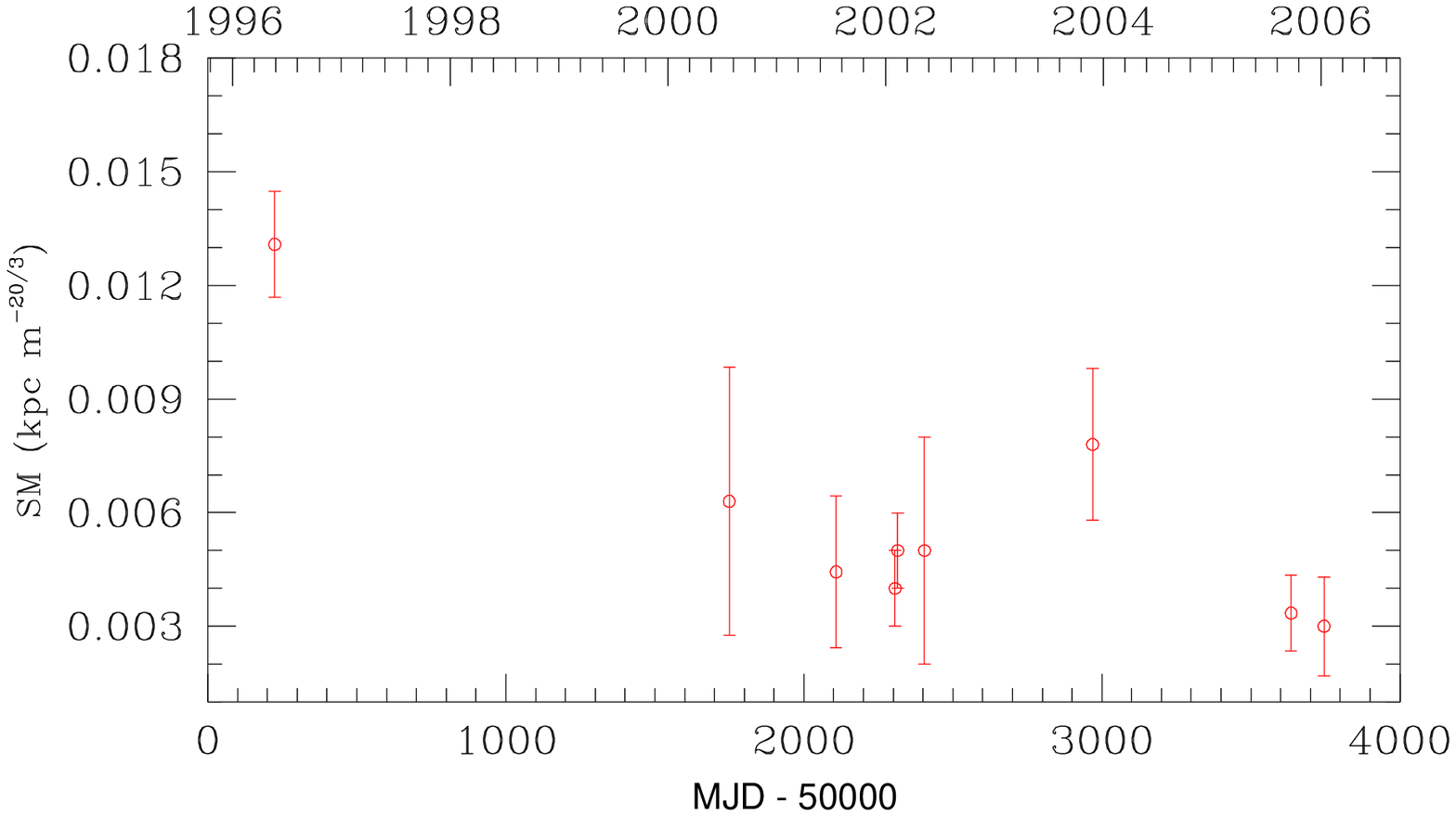}
\plotone{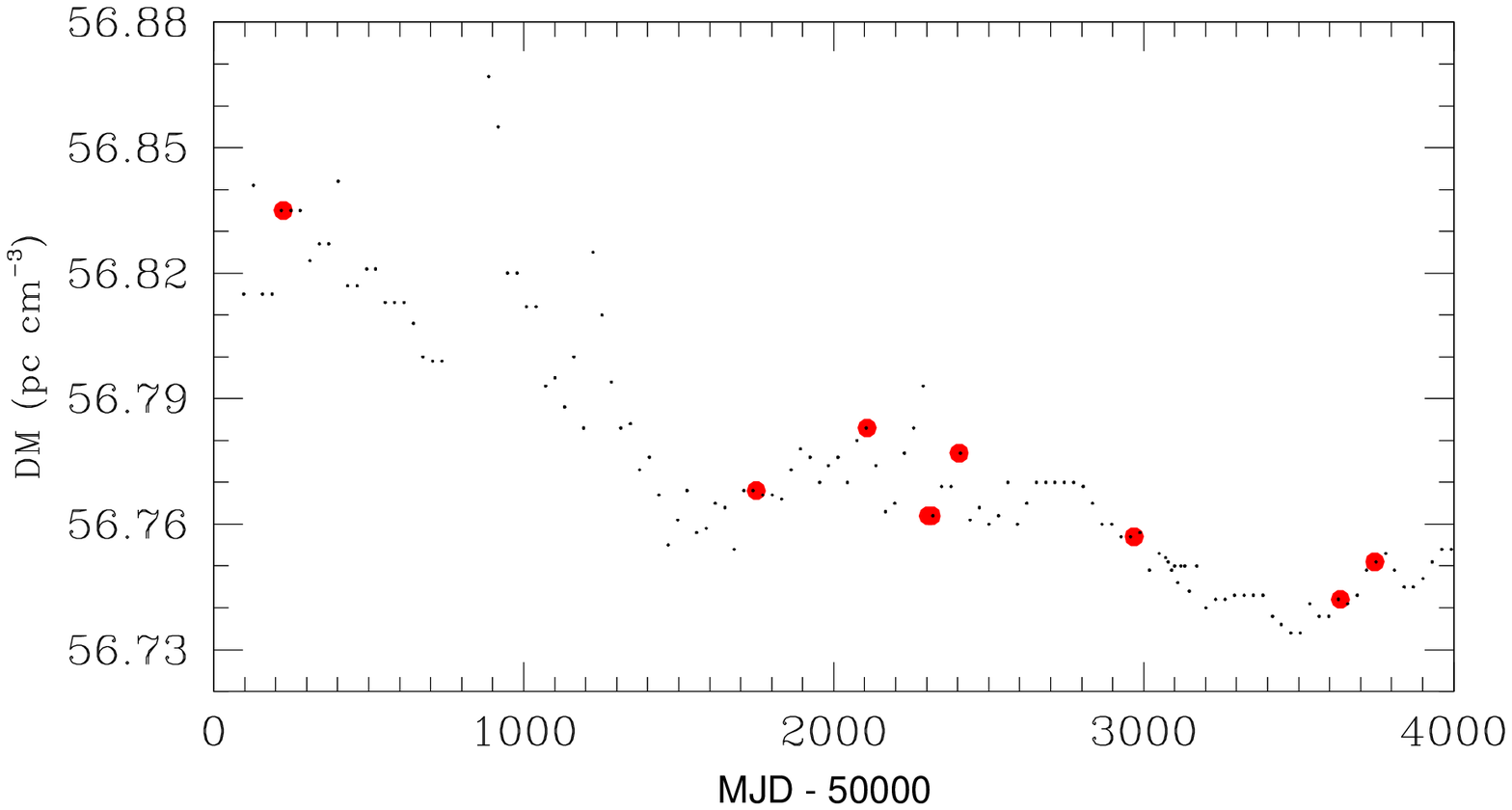}
\caption{Measurements of DM and SM of the Crab pulsar from observations over 
the past 10 years (1996 to 2006). 
The SM values ({\it upper panel}) are estimated from the published 
pulse-broadening measurements and decorrelation bandwidths --
1996.4: \citet{sallmenetal1999};
2000.6: \citet{kostyuketal2003};
2001.5: \citet{kuzminetal2002};
2002.1, 2002.4: \citet{cordesetal2004};
2003.9: \citet{popovetal2006};
2005.6: \citet{bhatetal2007};
2006.0: our observations. 
The DM measurements are from the Jodrell Bank monthly ephemeris
({\it lower panel}); the thick solid circles correspond to the 
observing epochs of the SM measurements.
}
\label{fig:dmsm}
\end{figure}

A closer look at the JB ephemeris reveals quite a systematic decrease in the 
Crab's DM variations between 1996 and 2006, with the lowest DM recorded 
near mid-2005 and a reversal of the trend in early 2006. 
A plot of SM estimates for the available \taud measurements during this period is shown 
in Fig.~\ref{fig:dmsm} along with the DM measurements at relevant epochs. 
These plots illustrate a gradual reduction in the Crab's DM and scattering over this
10-yr time span. 
Such variations are too large and smooth to be caused by refractive scintillation effects 
in the ISM, and are therefore indicative of a nebular origin. 

The material within the Crab Nebula, specifically the perturbed thermal plasma associated 
with it, has often been advanced as the source of excessive scattering and anomalous DM 
variations on several occasions. 
The most remarkable observations in support of this are the anomalous scattering recorded in 
1974-1975 \citep{lyne-thorne75,isaac-rankin77} and the reflection event 
in 1997 \citep{backeretal2000,lyneetal2001}. 
The scattering event in 1974 was especially noted for its extreme activity, where the pulsar's 
DM rose by 0.07 \dmu and the scattering increased by an order an magnitude over a time span of 
several months.
\citet{isaac-rankin77} ascribed this to a two-component scattering-screen model where the 
variable screen associated with the nebula gives rise to such rapid changes in both DM and 
scattering. 
The anomalous dispersion event in 1997 was seen as discrete moving echoes of the pulse, and was 
interpreted as reflections from an ionized shell in the outer parts of the nebula 
by \citet{lyneetal2001}, and in terms of variable optics of a triangular prism located 
in the interface of the nebula and the supernova ejecta by \citet{backeretal2000}. 

Fig.~\ref{fig:dmsm} reveals a similar change in DM and a large but less dramatic change in scattering. 
A direct interpretation of such observations can be made along the lines of a variable scattering 
screen model, as we discuss in detail in the following section.

\subsection{Implications for the Nebular Structure and Densities} \label{s:nebula}

The observed changes in DM and SM can be used to constrain the combination of the electron density and the ``fluctuation parameter'' in the nebular region, denoted as \nec and \fc respectively. 
Following \citet{cordes-lazio2003} and \citet{bhatetal2004}, the measured decrements in DM and SM can be expressed as
$\deldm = \nec \dels$ and $\delsm = \cnc \dels$, where \dels is the size of the nebular scattering region and \cnc is the equivalent turbulent intensity (assuming a uniform distribution of scattering material). 
The parameters \fc and \cnc can be related as \citep{taylor-cordes1993,cordes-lazio2002} 
$\cnc = \csm \fc \necsq $, where $\csm = [ 3 (2 \pi)^{1/3} ]^{-1} \, \ku $ for a Kolmogorov spectrum, and 
$\ku$ =10.2 $\kuu$ to yield SM in units of $\smu $. 
The above expressions can be combined to yield the ratio of \delsm and \deldm, and is given by
\be
{ \delsm \over \deldm } = \csm \fc \nec
\ee
Thus our measurements of \delsm = 0.01 $\smu$ and \deldm = 0.09 \dmu yield \fc \nec = 55.
The fluctuation parameter is essentially a product of normalized variances (at small and large scales) and other terms such as the outer scale and filling factor. 
The electron density \nec is unknown, but assuming a nominal value of 1 \neu, we get \fc = 55, which is much larger than that typical of the Galactic spiral arms ($F \sim 10$).

The large values of \delsm and \fc indeed confirm the material within the nebula as the source of 
excessively strong scattering.
While the scattering and dispersion events of 1974 and 1997 were interpreted in terms of a single 
large structure with electron density $\sim$1,500 \neu, the long-term systematic variations in 
dispersion and scattering as shown in Fig.~\ref{fig:dmsm} can be interpreted in terms of a scenario 
whereby the nebular segment of the LOS is populated by many smaller structures of much lower densities. 
A variation in the number density of such structures may then account for the observed changes in DM and SM. 
As the nebula is thought to comprise fine structure on many length scales, perhaps even on scales much finer than the filamentary structure suggested by optical observations \citep[e.g.][]{hesteretal1995}, such a picture seems quite plausible.

For the sake of simplicity, if we model the measured changes in scattering and dispersion to 
arise from N such structures of size \delsc and density \nesc, the resultant contributions to 
SM and DM are given by
\be
\delsm \equiv \sum _{i=1} ^{i=N} \smsci = N \, \csm \fcs \nessq \delsc
\ee
\be
\deldm \equiv \sum _{i=1} ^{i=N} \dmsci = N \, \nesc \delsc
\ee
Following equations (4)--(6) and the constraint $\fc \nec = 55$, the electron density of such a structure 
can be estimated as
\be
\nesc = \left( { \delsm \over 55 \, \csm } \right) \left( { 1 \over N \, \delsc } \right)
\ee
Thus, with just two direct measurements alone (\deldm and \delsm), it is hard to constrain 
all 3 free parameters of the model. 
However, assuming a reasonable value for \delsc $\sim$ $10^{-5}$ pc (i.e. an order of magnitude smaller 
than that implied by the 1997 reflection event), we estimate \nesc $\sim$ 100 \neu for N$\sim$100. 
This is almost an order of magnitude smaller than the densities required to produce the reflection 
event.
Indeed, several different combinations of size and density are possible; nonetheless, the underlying 
picture is the presence of many moderately-dense structures in the nebular region, with a 
filling factor $\sim$ $10^{-3}$.

We note that the electron density estimated above, estimates of the electron temperature 
in the Crab nebula \citep{temimetal2006,hesteretal1995,davidson1979} that yield values 
between 6000 and 16000 K, and the size of the Crab nebula 
(approximately $6^{\prime}$ = 3 pc, for a distance of 2 kpc), 
imply that the nebula should produce a minimal optical depth to free-free
absorption at GHz frequencies and a significant optical depth at lower
frequencies.  We estimate, based on these parameters, that a free-free
optical depth ($\tau _{ff}$) of 0.007 at 1 GHz may be possible, along
lines of sight that include the scattering structures discussed above.
For radio sources behind the Crab nebula, this would give a decrease
in the observed flux density, compared to the intrinsic flux density,
of 0.7\% due to free-free absorption at this frequency.  At lower
frequencies this decrease would be more significant: 3\% at 500 MHz,
12\% at 250 MHz, and 60\% at 100 MHz.

If a radio interferometer operating at these frequencies could achieve
an angular resolution high enough to resolve the Crab nebula and
detect compact sources of emission behind the nebula, it may be
possible to survey the free-free absorption due to the nebula along
many lines of sight, probing the structures that are producing the
scattering of the pulsar emission.  The upcoming future instruments
such as the extended LOFAR telescope in The Netherlands and Europe, 
or the Murchison Wide-field Array (MWA) in Western Australia may be 
able to undertake such a survey.

\section{Summary and Conclusions} \label{s:conc}

Using the ATCA and a baseband recording system, we detected more than 700 giant pulses 
from the Crab pulsar from our continuous and uniform recording at 1300 and 1470 MHz 
over 3 hours. 
This large sample is used for investigating statistical properties of giant pulses,
such as their amplitude, width, arrival time and energy distributions.
The amplitude distribution follows roughly a power-law with a slope of $-2.33\pm0.15$,
which is shallower compared to those from previous observations at lower frequencies.
The pulse widths show an exponential-tailed distribution and there is a tendency for 
stronger pulses to be narrower.
A majority of pulses (87\%) tend to occur within a narrow phase window ($\pm 200$ $\mu$s) 
of the main-pulse region.
Finally, the distribution of pulse energies follows a power-law with a slope of $-1.6$, and 
there is evidence for a break near $\sim$10 kJy $\mu$s.

The brightest pulse detected in our data has a peak amplitude of 45 kJy and a width of
0.5 $\mu$s, implying a brightness temperature of $ 10^{35}$ K, which makes it the 
brightest pulse recorded from the Crab pulsar at the L-band frequencies (1--2 GHz). 

Our observations show that many giant pulses comprise multiple narrow components at a 
time resolution $\sim$128 ns, which confirms the fundamental picture of giant pulses being 
superpositions of extremely narrow bursts.
Further, the measured pulse shape is marked by an unusually low degree of scattering,
with a pulse-broadening time of $ 0.8 \pm 0.4 $ $\mu$s that is the lowest estimated yet
towards the Crab from observations so far. 
Further, the pulsar's DM is determined to be $ 56.751 \pm 0.001 $ \dmu, which is significantly
lower than those measured near the epochs of previous scattering measurements.

Our measurements of DM and scattering, together with published data and the Jodrell Bank 
monthly ephemeris, unveil a systematic and slow decrease in the Crab's DM and scattering 
over the past 10 yr.
These variations are too large and smooth to be caused by the intervening ISM but can be 
attributed to the material within the nebula.
Our analysis hints at there being large-scale inhomogeneities in the distribution of small-scale 
density structures in the nebular region, with a plausible interpretation involving many ($\sim$
100) dense ($\sim$100 \neu) structures. 
A variation in their number density or size can potentially lead to the observed
changes in DM and scattering.
Such a possibility can be further investigated by obtaining independent constraints on the nebular 
electron densities (e.g. via free-free absorptions at low radio frequencies) and through 
future observations to monitor the pulsar's DM and scattering.

\medskip
\noindent
{\it Acknowledgements:} The ATCA is part of the Australia Telescope, which is funded
by CSIRO for operation as a National Facility by ATNF. Data processing was carried 
out at Swinburne University's Supercomputing Facility. We thank Matthew Bailes and 
Simon Johnston for fruitful discussions, and Willem van Straten and Joris Verbiest
for a critical reading of the manuscript. 
We also thank the referee for a critical review and several inspiring comments and 
suggestions that helped improve the presentation and clarity of the paper.
This work is supported by the MNRF research grant to Swinburne University of Technology.


\end{document}